# The Search for Superheavy Elements: Historical and Philosophical Perspectives

Helge Kragh*

**Abstract**: The heaviest of the transuranic elements known as superheavy elements (SHE) are produced in nuclear reactions in a few specialized laboratories located in Germany, the U.S., Russia, and Japan. The history of this branch of physical science provides several case studies of interest to the philosophy and sociology of modern science. The story of SHE illuminates the crucial notion of what constitutes a chemical element, what the criteria are for discovering a new element, and how an element is assigned a name. The story also cast light on the sometimes uneasy relationship between physics and chemistry. It is far from obvious that elements with $Z > 110$ exist in the same sense as oxygen or sodium exists. The answers are not given by nature but by international commissions responsible for the criteria and evaluation of discovery claims. The works of these commissions and of SHE research in general have often been controversial.

## 1. Introduction

Beyond uranium the periodic table includes no less than 26 elements which have all been manufactured in the laboratory and the best known of which is plutonium of atomic number 94. The heaviest of the transuranic elements are often called "superheavy elements" (SHE), a term with no precise meaning but which often refers to the transactinide elements with $Z$ ranging from 103 to 120. In some cases the term is used only for $Z > 110$. So far the last of the superheavy elements is $Z = 118$, a substance which received official recognition as an element in 2016 and is named oganesson, chemical symbol Og. The term superheavy element (SHE) owes its origin to John Wheeler, who in the 1950s examined theoretically the limits of nuclear stability. He suggested that atomic nuclei twice as heavy as the known nuclei might be ascribed "experimental testable reality" in the sense of having a lifetime greater than $10^{-4}$ second.[1]

---

* Niels Bohr Institute, University of Copenhagen, Denmark. E-mail: helge.kragh@nbi.ku.dk.
[1] Wheeler (1955), p. 181; Werner and Wheeler (1958).

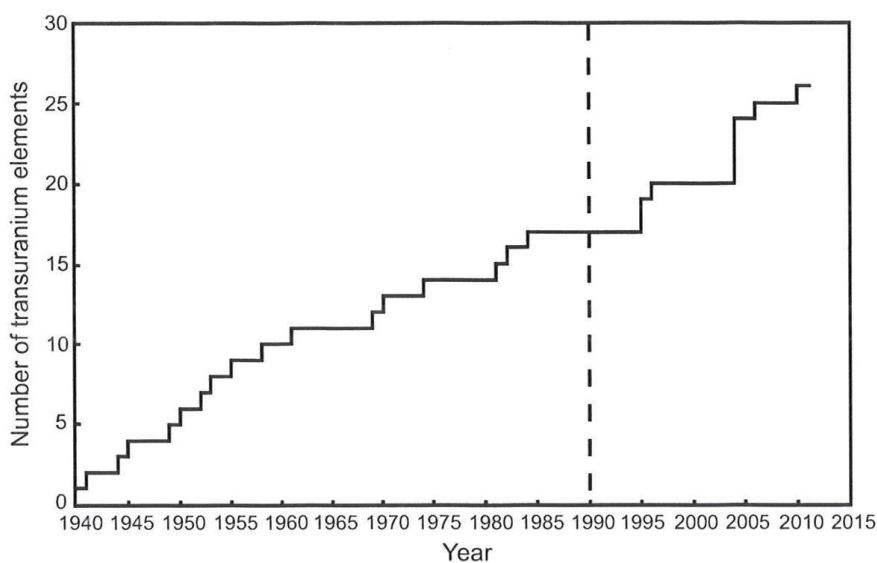

Figure 1. Number of discovered transuranic elements. Source: Thoenessen (2013), p. 9.

A decade later experimentalists began looking for SHE produced in artificial nuclear reactions. Early researchers sometimes defined SHE as elements located in or near the theoretically predicted "island of stability" around $(Z, N) \cong (114, 184)$. Thus, Glenn Seaborg and Walter Loveland suggested in 1990 that the term SHE should be associated with "an element whose lifetime is strikingly longer than its neighbors in the chart of the nuclides."[2] The first scientific paper with "superheavy elements" in the title appeared in 1966 and fifty years later the cumulative number of such papers was 702.

Theories and experiments on SHE have attracted much scientific interest for more than half a century, but only little has been written on the subject from a historical perspective and even less from a philosophical perspective.[3] And yet the subject is of considerable interest from the point of view of history, philosophy and sociology of science. For one thing, SHE research offers a new window to the current chemistry-physics relationship. For another thing, it provides case studies illuminating key concepts such as prediction, discovery, and reproducibility of experiments; and, related to sociology of science, fraud and scientific misconduct. The SHE story also problematizes the crucial notion of what constitutes a chemical element. Can one reasonably claim that superheavy nuclei exist in the same sense that the element oxygen exists? Apart from this ontological question, SHE research involves the epistemic question of how knowledge of a new SHE is obtained and what the criteria for accepting discovery claims are. And, what is equally important, *who* are responsible for the criteria and for evaluating discovery claims?

---

[2] Seaborg and Loveland (1990), p. 287. The same terminology was used in Thompson and Tsang (1972), one of the earliest SHE review articles.

[3] There is no comprehensive history of modern SHE research. Weeks and Leicester (1968), pp. 840-857, covers elements from $Z = 93$ to 103. For more scientifically oriented histories, see Seaborg and Loveland (1990) and Hofmann (2002).

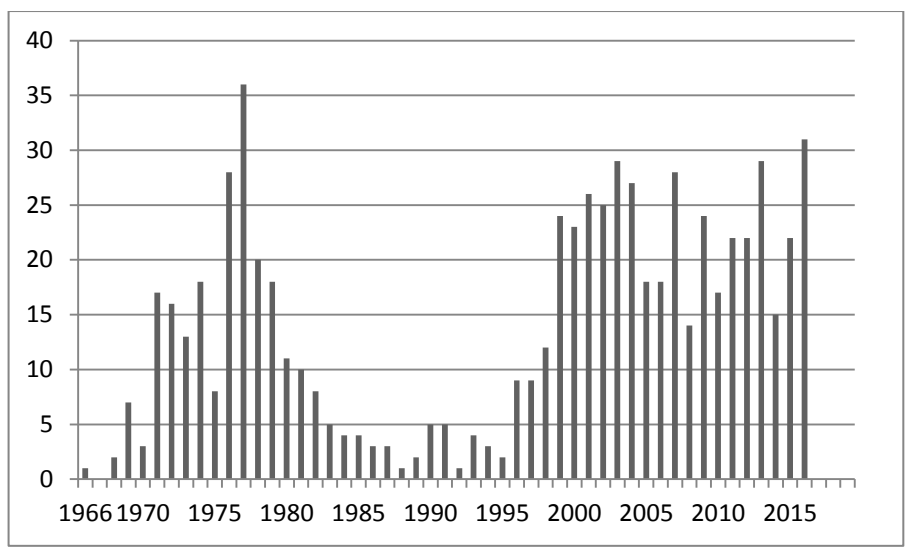

Figure 2. Number of scientific papers including the term "superheavy elements" in the title. Source: Web of Science.

## 2. Transuranic alchemy

Ever since the late nineteenth century a few chemists speculated of a possible end of the periodic table and the possibility of natural transuranic elements. With the advent of the Bohr-Sommerfeld semi-classical atomic theory these speculations were turned into calculations of the maximum number of atomic electrons. Niels Bohr, Richard Swinne, Walther Kossel and other atomic physicists suggested electron configurations for several elements beyond uranium. According to Bohr, the electrons in $Z$ = 118 were arranged as (2, 8, 18, 32, 32, 18, 8). Interesting as these early ideas are, they remained hypothetical and consequently attracted but little scientific attention. They will not be covered in this essay.[4]

---

[4] See Kragh (2013) for a review of early work on transuranic elements and attempts to predict their chemical properties.

## 2.1. From Fermi to Seaborg

The possibility of manufacturing transuranic elements in the laboratory only became a possibility after 1932, when the neutron was discovered and the atomic nucleus reconceptualised as a proton-neutron rather than a proton-electron composite. Enrico Fermi and his group in Rome, including Edoardo Amaldi, Emilio Segré and Oscar D'Agostino, studied systematically neutron reactions with all the elements of the periodic system, paying particular attention to the heaviest elements. Famously, when they bombarded uranium with slow neutrons they obtained results which suggested that they had obtained an element of atomic number greater than 92. Indeed, for a time Fermi and his group believed that they had produced elements with $Z = 93$ and $Z = 94$. When Fermi was awarded the 1938 Nobel Prize in physics it was in part because he had "succeeded in producing two new elements, 93 and 94 in rank number. These new elements he called Ausenium and Hesperium."[5] The two elements were believed to have been formed as

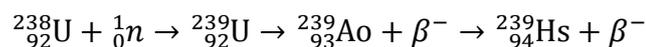

$$^{238}_{92}U + ^{1}_{0}n \rightarrow ^{239}_{92}U \rightarrow ^{239}_{93}Ao + \beta^- \rightarrow ^{239}_{94}Hs + \beta^-$$

However, it quickly turned out that the Nobel Foundation's support of ausenium and hesperium was an embarrassing mistake. The physics prize of the following year was awarded to Ernest Lawrence for his invention of the cyclotron, and also in this case did the prize motivation refer to the "artificial radioactive elements" resulting from the invention.

Although this essay is about artificially made transuranic elements it is worth recalling that the first new element ever produced in the laboratory was sub-uranic. In 1937 Segré and his collaborator, the Italian mineralogist Carlo Perrier, analyzed plates of molybdenum irradiated with deuterons and neutrons from the Berkeley cyclotron. They were able to identify two isotopes of element 43, for which they proposed the name technetium ten years later. There had earlier been several unconfirmed claims of having detected the element in natural sources, but Segré and Perrier are recognized as discoverers.[6] Segré is also recognized as the co-discoverer, together with Dale Corson and Kenneth MacKenzie, of element 85 which was produced in Berkeley in 1940 by bombarding Bi-209 with alpha particles. In 1947 they suggested the name astatine for it. Tiny amounts of astatine exist in nature, and there were previous claims of having identified the very rare element.[7]

---

[5] Presentation speech of Henning Pleiel of 10 December 1938. See https://www.nobelprize.org/nobel_prizes/physics/laureates/1938/press.html. Fermi was nominated for both the physics and chemistry prize.

[6] See Scerri (2013), pp. 116-143 for the complicated discovery history of technetium, which for a time was also known under the names "masurium" or "illinium."

The early attempts to produce transuranic elements are thoroughly described in the literature, not least because of their intimate connection to the discovery of the fission of the uranium nucleus.[8] We only need to point out that it was investigations of fission fragments that led Edwin McMillan and Philip Abelson to conclude that they had detected element 93 as a decay product of neutron-produced U-239. This first transuranic element was announced in *Physical Review* on 27 May 1940, at a time when the self-imposed publication ban caused by the war was still not effective. The element was called neptunium, chemical symbol Np. Even more importantly, in 1941 Seaborg and collaborators prepared the isotope Np-238 which beta-decayed to a daughter product which they identified as an element of $Z = 94$. They also produced the more important isotope of mass number 239, soon showed to be fissionable. The discovery paper was submitted on 7 March 1941, but as a result of the war it only appeared in print in April 1946.[9]

The name of the new element, plutonium (Pu), first appeared in a paper of 1948. Following the names of uranium and neptunium, it was named after the next planet, Pluto (which today is no longer classified as a planet). Needless to say, it quickly turned out that plutonium was more than just another exotic element of interest to the scientists only. The inhabitants of Nagasaki experienced that on 9 July 1945. Several of the transuranic elements have been made in visible quantities and a few of them, such as long-lived isotopes of curium and americium, have even found applications in science and industry. Plutonium is unique by being the only synthetic element produced in very large quantities. It is estimated that today the world stockpile of the element is about 500 tons.[10] The long half-life of plutonium ($2.4 \times 10^4$ years for Pu-239) means that the element is not just an ephemeral visitor on Earth but will remain with us for thousands of years to come.

---

[7] The discovery history of element 85 is described in Thornton and Burdette (2010).
[8] See, for example, Sime (2000) and Seaborg (2002). For a detailed contemporaneous review of the confused situation shortly before the discovery of fission, see Quill (1938).
[9] Seaborg, Wahl and Kennedy (1946). Together with other of Seaborg's papers it is reprinted in Seaborg (1994a).
[10] https://en.wikipedia.org/wiki/Plutonium

The early history of transuranic elements was completely dominated by a group of Californian chemists and physicists led by Seaborg and Albert Ghiorso. Elements 95 and 96 were first identified in 1944 at the Metallurgical Laboratory in Chicago and named americium (Am) and curium (Cm), respectively.[11] Whereas the first was identified as a decay product of Pu-241, the latter was produced by bombarding Pu-239 with alpha particles in the Berkeley cyclotron. It was problems with synthesizing curium that led Seaborg to propose, first in secret report of 1944, a modified form of the periodic table with what he called an actinide series with elements from $Z$ = 89 to 103. The first version of the periodic table including Seaborg's actinide series appeared in *Chemical & Engineering News* at the end of 1945.[12]

After the war followed the discovery of $Z$ = 97 (berkelium, Bk) and $Z$ = 98 (californium, Cf) which were announced in 1950 and also obtained by means of a beam of cyclotron-produced alpha particles. The identification of Cf was accomplished with a total of some 5000 atoms with a half-life of 44 min. Although this was a very small amount, it was much greater than in future element discoveries. In 1951, at a time when six transuranic elements had been added to the periodic system, Seaborg and McMillan were awarded the Nobel Prize in chemistry "for their discoveries in the chemistry of the transuranium elements." Incidentally, the Seaborg-McMillan prize of 1951 is the only Nobel Prize awarded to research in transuranic or superheavy elements. Notice that the prize was in chemistry and not physics, indicating that discoveries of new elements traditionally belong to the first science. While Seaborg was a chemist, MacMillan was a physicist. As MacMillan pointed out in his Nobel lecture, "in spite of what the Nobel Prize Committee may think, I am not a chemist."[13]

Elements 99 and 100, named einsteinium (Es) and fermium (Fm), were first identified in late 1952, not in a planned experiment but in the fallout from a test of the American hydrogen bomb. The discovery team led by Ghiorso only published its findings in 1955, a delay caused by orders from the U.S. military. The priority of the

---

[11] See Seaborg (1994b) for a lively account of the two elements' naming history, and Kostecka (2008) for the discovery and applications of americium.

[12] Seaborg (1995). The 1944 memorandum is reprinted in Seaborg (1994a), pp. 145-148. Seaborg later introduced also the transactinide series ($Z$ from 104 to 121) and even the superactinide series ($Z$ from 122 to 153).

[13] MacMillan (1951). MacMillan was far from the only physicist to be awarded a Nobel Prize in chemistry. See the table in Kragh (1999), p. 432.

American team was generally recognized and that although the synthesis of an isotope of element 100 was reported in 1954 by Swedish physicists. The Ghiorso team suggested the symbol E for einsteinium, but it was later changed to Es. Also in 1955, the discovery of element 101 (mendelevium, Md) was announced by the Berkeley group using its cyclotron to irradiate a tiny sample of Es-253 with alpha particles:

$$^{4}_{2}\text{He} + {}^{253}_{99}\text{Es} \rightarrow {}^{256}_{101}\text{Md} + {}^{1}_{0}n$$

The delicate experiment led to the production of only 17 atoms of Md-256 identified one by one, but that was enough to discover the new element. For the discovery story of element 102, named nobelium, see Section 3.4.

It took several more years until the last element in the actinide group, lawrencium of atomic number 103 was discovered, this time in a more involved process. What appeared to be solid evidence for $Z$ = 103 was obtained in Berkeley in 1961 with the production of the Lr-258 isotope, but at the time Berkeley no longer had monopoly in the manufacture of SHE. Russian scientists at the Joint Institute for Nuclear Research (JINR) founded in Dubna in 1956 argued that the American results were not fully convincing. Later in the 1960s they produced their own isotope of the element, reportedly Lr-256, which they proposed to call rutherfordium. The doubts concerning the discovery were finally dispelled, but only in 1992 did a joint working group established by the International Union of Pure and Applied Chemistry (IUPAC) and the International Union of Pure and Applied Physics (IUPAP) accept the element. The discovery was assigned jointly to Berkeley and Dubna (see also Section 4.2). Although the name lawrencium (after Ernest Lawrence, the inventor of the cyclotron) was retained, in 1997 its chemical symbol was changed from Lw to Lr. At the same occasion the symbol of einsteinium was changed from E to Es and the one of mendelevium from Mv to Md.

## 2.2. Aspects of SHE research

Much of the impetus for SHE research derives from theories of nuclear structure and in particular from predictions based on the shell model developed in the late 1940s independently by Maria Goeppert Mayer in the U.S. and Hans Jensen and collaborators in Germany. The idea of a nuclear shell structure can be found earlier, first in a 1933 paper by the German physicists Walter Elsasser and Kurt Guggenheimer. According to Mayer, nuclei with 2, 8, 20, 50, 82, and 126 protons or neutrons were particularly stable. These were "magical numbers" representing closed shells in the nucleus, an idea which mineralogists had anticipated much earlier.[14] The theoretical possibility of a relatively stable element of $Z = 126$ seemed remote from laboratory physics, but in the late 1960s more sophisticated nuclear models indicated that $Z = 114$ rather than $Z = 126$ was a magic number. The region around $(Z, N) = (114, 184)$ – a "doubly magical" nucleus – was expected to represent nuclei with a relatively long half-life and therefore accessible to experimental study. The region became known as an "island of stability," a term that may first have appeared in the physics literature in 1966.[15] The hope of the experimenters was to reach the fabled island, if it existed.

Although all known elements of $Z > 92$ are produced in the laboratory, there is the faint possibility that some of them may exist naturally, either on the Earth, in cosmic rays or perhaps in the interior of stars. Such speculations were entertained as early as the 1920s, many years before the first synthesis of transuranic elements. With the idea of an island of stability of mass number $A \sim 300$ including nuclides with half-lives as long as $10^8$ years or even longer, interest in the question of naturally occurring SHE took a new and less speculative turn. From about 1970 several researchers looked for evidence of SHE in cosmic rays, meteorites, or terrestrial ores. For example, in 1973 a team of American scientists reported an extensive investigation of SHE with $Z$ from 110 to 119 in terrestrial and meteoric minerals, including among their samples 60-million-old sharks' teeth.[16] The result of their search was negative.

Other scientists examined the possibility of stellar nucleosynthesis of SHE. Given that absorption lines of technetium had been observed in the spectra of red giant stars since 1952, it was not unreasonable to assume that also SHE might be produced by stellar nucleosynthesis.[17] According to the American astrophysicists David Schramm and William Fowler, SHE might indeed be produced in explosive stellar events.[18] However, other calculations led to more pessimistic results and since no evidence of naturally occurring SHE was found, scientists came to agree that SHE do not exist in nature. Superheavy elements seem to be the exclusive business of nuclear laboratories. Claims of having detected the elements in ordinary matter are met with more than just ordinary scepticism (see Section 3.2), and yet a few scientists are still pursuing the search for SHE in nature.

---

[14] The first anticipation of magical numbers was due to the Swiss mineralogist Paul Niggli in 1921. See Kragh (2000a) and, for the development of the shell model, Mladenović (1998), pp. 287-305.

[15] Myers and Swiatecki (1966).

[16] Stoughton et al. (1973). For other early searches for natural SHE, see Thompson and Tsang (1972).

[17] Merrill (1952).

[18] Schramm and Fowler (1971).

Most scientists involved in SHE research consider it a field of fundamental science unrelated to applications. In a recent interview Yuri Oganessian justified his research field by saying that "it is about tackling fundamental questions in atomic physics." Foremost among the questions is the prediction of an island of stability. According to Oganessian: "Theorists predict that there should be some superheavy atoms, with certain combinations of protons and neutrons, that are extremely stable." Referring to elements heavier than $Z$ = 112 he continued:

> Their lifetimes are extremely small, but if neutrons are added to the nuclei of these atoms, their lifetime grows. Adding eight neutrons to the heaviest known isotopes of elements 110, 111, 112 and even 113 increases their lifetime by around 100,000 times … but we are still far from the top of the island where atoms may have lifetimes of perhaps millions of years. We will need new machines to reach it.[19]

Other leading SHE scientists have expressed a similar *l'art pour l'art* attitude. Sigurd Hofmann refers to the "sense of the excitement which has motivated workers in this field" and suggests that the motivation for study SHE is "because we are curious."[20] But of course one may always fall back on the mantra, as two SHE nuclear chemists did in 1972, that "Practical and useful applications would be forthcoming eventually, as is always the case with basic research."[21]

## 3. Controversial discoveries and discovery claims

### 3.1. General considerations

Many of the discoveries of transuranic elements and the naming of them have been surrounded by controversies of priority often involving national and political elements. This is hardly surprising, given that as a rule discoveries of new elements occurring in nature have also been followed by controversies of this kind. Early examples are oxygen and vanadium, and later examples are hafnium and rhenium.[22] While most of the priority conflicts have been contemporaneous with the discovery claims, in some cases they have occurred retrospectively and been due to later intervention of scientists or historians of science. But of course priority disputes are not limited to chemical elements as they constitute a general feature in science. As pointed out by the sociologist Robert Merton in a classical study, priority conflicts are part and parcel of the development of science and not anomalous features.[23]

---

[19] Interview in Gray (2017).
[20] Hofmann (2002), p. 205.
[21] Thompson and Tsang (1972), p. 1055.

Naturally, a priority conflict concerning a new element can only be resolved on the basis of some commonly agreed criteria of what constitutes a discovery. According to E. Rancke-Madsen, a Danish chemist and historian of chemistry, two conditions are necessary and sufficient for a person to be accepted as the discoverer of an element:

> 1. He has observed the existence of a new substance which is different from earlier described substances, and this new substance is recognized by him or later by scientists as being elemental. 2. He must have published the discovery of the new substance in such a manner that it has been noticed by contemporaries outside the immediate circle of the discoverer, usually in a periodical or monograph.[24]

These conditions are reasonable, but they fail to discriminate between discovery claims and discoveries, or between claimed observations and scientifically valid observations. Moreover, they do not take into account that priority disputes are settled by negotiations according to social norms within the relevant scientific community and that these norms may change over time.

Since the early twentieth century the recognition and naming of new elements has been formalized and institutionalized. In the earlier period the unwritten rule was that the discoverer named the element, although there was no agreed system and in many cases also no agreement as to who the discovery should be credited.[25]

The first national atomic weight committees appeared in the 1890s, starting with a committee appointed by the American Chemical Society (ACS) in 1892.

---

[22] The controversy over $Z = 72$ is analysed in Kragh (1980) and the case of $Z = 43$ in Zingales (2005). For some other cases, see Scerri (2013).
[23] Merton (1973), first published 1957.
[24] Rancke-Madsen (1976); Scerri (2013), pp. xxv-xxiii.
[25] See Childs (1998) for the naming of elements through history.

Together with this committee, the most important of the national committees was the German Atomic Weight Commission formed in 1897. In 1899 the International Committee on Atomic Weights (ICAW) was established, consisting of 56 delegates from eleven countries.[26] After World War I the committee was reorganized within the structure of IUPAC which was founded in 1919 as part of IRC, the International Research Council. However, initially only the chemical societies from the allied powers and neutral countries were admitted to IUPAC. In connection with the German membership in 1930 IUPAC was renamed IUC (International Union of Chemistry), but in 1947 the organization returned to its earlier name.[27] Much later, in 1979, the atomic weight committee's name was changed to the Commission on Atomic Weights and Isotopic Abundances (which again, in 2002, was changed to the Commission on Isotopic Abundances and Atomic Weights, CIAAW).

The names of new elements were not officially approved by ICAW except that they entered the committee's versions of the periodic table. In the late 1940s questions of names and priority were conferred to IUPAC's Commission on Nomenclature of Inorganic Chemistry (CNIC) which since then has played an important if not always fortunate role. CNIC was terminated in 2002 and its responsibilities transferred to the Inorganic Chemistry Division under IUPAC.

At a conference in London in 1947 it was agreed that only CNIC could recommend a name to the union and that the final decision remained with IUPAC's executive council. "It has been accepted in the past that the discoverers of a new element had the sole right to name it," it was stated. But IUPAC did not agree with the tradition:

> Priority is only one factor to be considered in deciding which is the best name for general international adoption. This presumptive right to name new elements is now accorded to the discoverers of new elements produced artificially, but subject to the approval of the Nomenclature Commission of IUPAC.[28]

The names of the transuranic elements were considered at the 15th IUC conference in 1949, taking place in Amsterdam. CNIC officially adopted the proposed names for elements 93 to 96 and also decided upon the names of some other elements.[29] For example, it was on this occasion that element 4 officially became beryllium and the older name glucinium was ruled out. It was also on this occasion that the name astatine was adopted for element 85, and technetium for element 43 at the expense of masurium and other names.

---

[26] See Holden (2004) for a historical review.
[27] See Fennell (1994) for the complex history of IUPAC.
[28] Quoted in Koppenol et al. (2002), p. 789.
[29] Anon. (1950); Coryell and Sugarman (1950); Koppenol (2005).

During the first decades of the twentieth century there were two basic criteria for recognizing the discovery of a new element, namely the optical spectrum and the atomic weight of the claimed element. On the other hand, recognition of the existence of a new element did not necessarily require that it was prepared in a pure state (one example is radium and an earlier one is fluorine). Since the mid-1920s the main criterion became the element's characteristic X-ray spectrum. For several of the superheavy elements none of these criteria are relevant since they have no definite atomic weight and also no spectrum based on electron transitions between different energy levels. Besides, X-ray spectroscopy requires amounts of matter much greater than the few atoms often produced in SHE reactions.

In 1978, after much delay and many years of preparation, CNIC suggested a naming procedure based directly on the atomic number and aimed specifically for the elements with $Z > 103$ which had not yet been discovered. The provisional names and corresponding symbols were based on Latin numbers, for instance unnilpentium (un-nil-pentium, Unp) for $Z = 105$ and ununbium (un-un-bium, Uub) for $Z = 112$. Contrary to the symbols of the known elements, the systematic symbols consisted of three letters. According to CNIC, "The systematic names and symbols for elements of atomic numbers greater than 103 are the only approved names and symbols for those elements until the approval of trivial names by IUPAC."[30] But although systematic and neutral, the unwieldy names were rarely used by the scientists who preferred to refer directly to the atomic numbers. As a linguist noted, the system was "a masterpiece of diplomacy but also of blandness and lack of imagination, as well as an etymological hodge-podge."[31] A search in Google Scholar shows that in the period 1980-1995 not a single research paper referred to unnilhexium, the systematic name for element 106 later to be named seaborgium.

To take care of priority questions IUPAC and IUPAP established in 1985 the Transfermium Working Group (TWG) consisting of nine scientists, two of which were nominated by IUPAC and seven by IUPAP. Members from USA, USSR and

---

[30] Orna (1982); Bera (1999).
[31] Diament (1991), p. 210.

West Germany were deliberately excluded to keep the group free of national bias, but of course TWG had consultations with SHE scientists from these countries. In end of the 1980s TWG visited the laboratories in Darmstadt, Berkeley and Dubna. The responsibility of the group, headed by the distinguished Oxford nuclear physicist Denys Wilkinson, was to formulate criteria for when an element was discovered and to evaluate discovery claims accordingly.[32] On the other hand, the naming of new elements was not part of the group's agenda. Wilkinson emphasized the neutrality and objectivity of the working group:

> The TWG consisted of nine men of goodwill who, conjointly and severally, spent some thousands of hours … in a microscopic and scrupulous analysis of the discovery of the transfermium elements. … We were utterly without bias, prejudice or pre-commitment and had no connection with any of the laboratories of chief concern; we did not care who had discovered the elements in question but agreed to find out.[33]

TWG held its first meeting in 1988 and published its criteria three years later in *Pure and Applied Chemistry*, since 1960 the official journal of IUPAC.[34] In 1991 TWG was disbanded and eight years later it was followed by another inter-union group of experts, called the Joint Working Party (JWP). This party or group consisted of only four members (an American, a Briton, a Japanese, and a Canadian), with two representing each of the participating scientific unions. Later again a new five-member joint working group was established in 2011 to examine claims of having discovered elements with $Z > 112$. The chair of this group was the American nuclear chemist Paul Karol, while the other four members were physicist. Thus, the former exclusion of members from claimant countries was no longer maintained.

### 3.2. Two failed discovery claims, *Z* = 112 and *Z* = 122

Many discovery claims end as just that – claims which remain unrecognized by the scientific community and therefore do not count as discoveries. In 1971 the Israeli physicist Amnon Marinov, together with five British physicists, reported their analysis of experiments done in CERN with high-energy protons hitting a target of tungsten. In their paper published in *Nature* they presented evidence for the production of element 112 by secondary reactions in the tungsten target. "We believe," they concluded, "that we may have observed the production of element 112."[35] The evidence was primarily in the form of observation and mass determinations of fission fragments combined with the prediction that $Z = 112$ was chemically homologous to mercury. In a later paper[36] Marinov interpreted the mass spectra in terms of reactions leading to nuclei with $N \cong 160$, such as

$$^{86}_{38}\text{Sr} + ^{186}_{74}\text{W} \rightarrow ^{272}_{112}\text{X}$$

---

[32] Fontani, Costa and Orna (2015), p. 386 states mistakenly that the TWG chairman was Geoffrey Wilkinson, the 1973 Nobel laureate in chemistry.
[33] Wilkinson et al. (1993b), p. 1824.
[34] Wapstra et al. (1991).

Marinov's group estimated the number of produced nuclei of mass number 112 to be about 500 and that they decayed by spontaneous fission with a half-life of a few weeks. Although the conclusion in the 1971 paper was cautious, Marinov had no doubt that he and his group had discovered a new SHE. In a later paper he spoke of "our claim for discovering element 112."[37] On the other hand, neither in 1971 nor in later publications did he or members of his group propose a name for the element.

Most physicists and chemists engaged in SHE research were unconvinced that element 112 had been discovered. In 1992 the TWG assessed the claim, concluding that "Data so far are insufficient to indicate that a new element has been produced."[38] Later assessments by the JWP did not change the verdict. Not only had independent attempts to duplicate the results failed, the JWP also argued that the interpretation of the Marinov collaboration in favour of $Z = 112$ rested on several assumptions of a speculative nature. The report was unequivocally critical:

> The collaboration's arguable use of forceful expressions such as "overwhelming evidence," "clear and proven," and "impossible to refute" is neither convincing nor swaying. Extraordinary intriguing phenomena, not much selective in their measured character, are, in part, necessary for the acceptance of the collaboration's interpretations of their data. The JWP needs much more to be able to relinquish its deeply felt unease that the tautological rationalization of the Marinov et al. measurements remains inadequate.[39]

---

[35] Marinov et al. (1971a); Marinov et al. (1971b). For Marinov's publications on SHE, see http://www.marinov-she-research.com/Long-Lived-Isomeric-States.html
[36] Marinov et al. (1984).
[37] Marinov et al. (2004).
[38] Barber et al. (1992), p. 507.
[39] Karol et al. (2003), p. 1606. An earlier and equally critical assessment appeared in Karol et al. (2001). JWP repeated the critique in Barber et al. (2009).

Marinov (meaning Marinov and his co-workers) did not accept the "harsh verdict" of the JWP, which he found was inconsistent with the earlier TWG assessment.[40] As documentation he did not refer to the TWG report but, unusually, to a private letter from one of the TWG members from which he quoted. His main objection against the negative conclusion of the JWP was based on the claimed discovery in the late 1990s of a new form of long-lived isomeric states of atomic nuclei. As Marinov saw it, this theoretical discovery justified the interpretation in terms of $Z = 112$ and he consequently maintained the discovery claim. But to no avail.[41] As the first JWP report made clear, this was not an acceptable justification:

> [Marinov's papers] continue to press arguments for the existence of very long-lived isomeric states of actinides and transactinides and of very high fusion cross-sections for their formation, each several orders of magnitude beyond current understanding. These extraordinary phenomena are, in part, necessary for the acceptance of the collaborations' interpretations. The JWP remained unmoved.[42]

Element 112 did become a reality, but in the version proposed by GSI (Gesellschaft für Schwerionenforschung) in Darmstadt, one of the centers of SHE research. In early 1996 a GSI team led by Sigurd Hofmann reported that two atoms of $Z = 112$ had been found.[43] The reaction was interpreted to be

$$^{208}_{82}\text{Pb} + ^{70}_{30}\text{Zn} \rightarrow ^{277}_{112}\text{Cn} + ^{1}_{0}n$$

Although Hofmann's group claimed to have produced and identified element 112 "unambiguously," the JWP concluded in its reports from 2001 and 2003 that the evidence for the new element was insufficient. It was only after new experiments and confirmations by teams in Dubna that the JWP was satisfied: "The JWP has agreed that the priority of the Hofmann et al. … collaborations' discovery of the element with atomic number 112 at GSI is acknowledged."[44] In agreement with the proposal of the discoverers the element was named copernicium, which was ratified by the IUPAC council the following year. The proposed chemical symbol was initially Cp, but it was soon realized that the symbol had earlier been applied for "cassiopeium" or what presently is lutetium ($Z$ = 71).[45] Consequently the symbol was changed to Cn.

---

[40] Marinov et al. (2004).
[41] Brandt (2005) supported the claim or rather considered it a possibility to be taken seriously. However, his support was not taken very seriously as he was a member of the Marinov collaboration.
[42] Karol et al. (2001), p. 964.
[43] Hofmann et al. (1996). The GSI authors did not refer to Marinov's earlier work.
[44] Barber et al. (2009), p. 1339; Bieber (2009).

The search for SHE is almost synonymous with attempts to synthesize very heavy atomic nuclei, but there is the slight possibility that trace amounts of SHE in unusual isomeric states may also occur in nature (Section 2.2). This is what Marinov hypothesized in papers of 2003 and 2007. The following year he and his team created attention by suggesting that they had detected a long-lived element with $Z$ = 122 in purified thorium. They summarized: "Mass spectral evidence has been obtained for the possible existence of a long-lived superheavy isotope with atomic mass number 292 and $t_{1/2} \geq 10^8$ y. Based on predicted chemical properties of element 122, it is probable that the isotope is $^{292}$122, but a somewhat higher $Z$ cannot absolutely be excluded."[46] The abundance of the new element was estimated to be about $10^{-12}$ relative to Th-232. Marinov made a similar claim of having detected small amounts of a long-lived state of $Z$ = 111 (roentgenium) inside a sample of gold.[47]

The claim of having found a monster nucleus in nature with $Z$ = 122 and $N$ = 170 did not meet with approval in the SHE community. Marinov had at the time a dubious reputation and his claim was scarcely taken seriously. Not only was the long half-life highly unusual and hard to explain theoretically, the experimental technique used for the identification was also criticized.[48] The JWP seems to have ignored the discovery claim of the Israeli maverick physicist, except that the group may have referred to it indirectly in a report of 2016.[49] In any case, still in his last paper, submitted to *Arxiv* in October 2011, Marinov defended the discovery of $Z$ = 122 and other SHE occurring in nature. He passed away two months later.

## 3.3. Another failed discovery claim, $Z$ = 118

Marinov's discovery claims of SHE were unfounded and based on insufficient data, but they were not fraudulent or involving scientific misconduct. In a field as competitive and potentially rewarding as SHE one should not be surprised that the latter category turns up, as it did in connection with an early discovery claim of $Z = 118$. In the summer of 1999 a group of 15 researchers at Lawrence Berkeley National Laboratory (LBNL), including the SHE veteran Albert Ghiorso, announced that it had detected two new elements, one with $Z = 118$ and the other its decay product with $Z = 116$. Bombarding a lead target with a beam of krypton nuclei, the group interpreted the results as due to fusion followed by an alpha decay:

$$^{208}_{82}\text{Pb} + ^{86}_{36}\text{Kr} \rightarrow ^{293}_{118}\text{X} + ^{1}_{0}n \quad \text{and} \quad ^{293}_{118}\text{X} \rightarrow ^{289}_{116}\text{Z} + ^{4}_{2}\text{He}$$

---

[45] Meija (2009). Cassiopeium, discovered by Auer von Welsbach in 1907, was for a period accepted by the German Atomic Weight Commission (but not by the International Commission) and it entered Niels Bohr's periodic table in the early 1920s.

[46] Marinov et al. (2010), p. 139.

[47] Marinov et al. (2011).

[48] Brumfiel (2008); Dean (2008).

[49] Öhrström and Reedijk (2016): "As serious claims associated with elements having Z = 119 or above have not yet been made, we note that, for the first time, the Periodic Table exists with all elements named and no proposed or pending new additions."

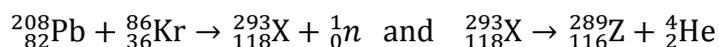

The lead author of the paper in *Physical Review Letters* was Viktor Ninov, a promising Bulgarian physicist who had previously worked at GSI in Darmstadt and recently been hired by LBNL. Ninov, who was in charge of the data analysis, argued that three instances of the reactions had been observed.[50] The news from Berkeley were exciting and at first generally accepted. However, what a first appeared to be a great success soon turned into something like a nightmare.

Neither the Darmstadt group nor other European laboratories were able to confirm the data on the reported synthesis of the nuclide of mass number 293. And when the Berkeley scientists tried to repeat and improve the experiments no sign of element 118 turned up. Eventually it dawned upon them that something was wrong. After several committees had examined the matter it was concluded that the records of 1999 were unreliable. Worse, some of the reported data were not real but had been fabricated, and Ninov seemed to be guilty of the fabrication. Although Ninov denied the charges, he was fired. "We retract our published claim for the synthesis of element 118," announced the 14 remaining authors in July 2002.[51] In emails to their colleagues in the SHE community, the leader of the Berkeley group apologized for the erroneous report and the consequences it might have caused.

What was immediately considered an embarrassing case of scientific misconduct attracted scientific as well as public interest.[52] The Ninov case was exposed at about the same time as the "Schön scandal," a reference to the German physicist J. Hendrik Schön who was fired from the Bell Labs for having manipulated and fabricated data in semiconductor research. The two cases did much to raise awareness of scientific misconduct in the physical sciences. As pointed out in *Chemical & Engineering News*, a weekly ACS journal, fraud of this kind should not happen, and yet it did – twice.[53]

---

[50] Ninov et al. (1999).

[51] *Physical Review Letters* **89** (2002): 039901. For an account of the scandal, see Johnson (2002). The CNIC report of 2003 briefly referred to the Berkeley claim and its retraction. Karol et al. (2003). See further Hofmann (2002), pp. 194-204.

[52] E.g., *Nature* **420** (2002): 728-729; *Science* **293** (2001): 777-778 and **297** (2002): 313-314; *New York Times*, 23 July 2002.

Not only had the claims of elements 116 and 118 evaporated, scientists at GSI were also forced to reconsider their own claim of having discovered elements 111 and 112. Evidence for these elements had been found in experiments from the mid-1990s in which Ninov, then at GSI, was responsible for the data analysis. According to a 2002 paper by Hofmann and collaborators, "In two cases … we found inconsistency of the data, which led to the conclusion, that for reasons not yet known to us, part of the data used for establishing these two [decay] chains were spuriously created."[54] Ten years later, looking back at the development, two GSI veterans still wrote as if the name of the forger was a mystery. But they also wrote that "The faking of results started at GSI and was exported to LBNL."[55] Fortunately, the basis for the German discovery claims of the two elements was not affected by the spurious data. The GSI priority to elements 111 and 112 was recognized by CNIC in 2003 and 2009 respectively.

And what about $Z = 118$? For how long would it remain ununoctium or eka-radon? A few atoms of mass number $A = 294$ were reported in 2002 and with greater confidence in 2006, by a collaboration led by the Russian physicist Yuri Oganessian, a highly esteemed specialist in SHE research. The big-science team consisted of 30 members, of whom 20 were from JINR and 10 from the Lawrence Livermore National Laboratory (LLNL).[56] The team observed three events from the fusion of Ca-48 and Cf-249, which they interpreted as

$$^{48}_{20}\text{Ca} + {}^{249}_{98}\text{Cf} \rightarrow {}^{294}_{118}\text{Z} + 3{}^{1}_{0}n$$

The formed nuclei were primarily identified by alpha decays with a half-life of only 0.9 milliseconds. In its 2011 report JWP concluded that the evidence did not yet satisfy the criteria for discovery. Consequently the results were confirmed and improved in subsequent experiments, but it took until 2016 before the JPW concluded that the work reported ten years earlier satisfied the criteria for discovery.[57] The provisional name ununoctium was now replaced with oganesson in recognition of the leader of the discovery team. The ending "-on" was in accordance with the IUPAC recommendations for elements in group 18 (except helium).[58]

---

[53] Jacoby (2002).
[54] Hofmann et al. (2002), p. 156.
[55] Armbruster and Münzenberg (2012), p. 298.
[56] Oganessian et al. (2006).

## 3.4. Nobelium, a controversial element

The first two decades of manufacturing new transuranic elements were dominated by Seaborg and his group of Californian physicists and nuclear chemists. Their dominance was first challenged in connection with element 102, which was announced in 1957 by a group of Swedish, American and British physicists working at Stockholm's Nobel Institute of Physics. A team of Swedish physicists had three years earlier used the institute's cyclotron to produce what they claimed was an isotope of $Z = 100$ but without determining its mass number.[59] In the 1957 experiment the extended group led a beam of C-13 ions to react with a sample of Cm-244 and based its discovery claim on the measured activity of an 8.5 MeV alpha emitter. The Stockholm physicists confidently announced not only a new transuranic element but also suggested "the name nobelium, symbol No, for the new element in recognition of Alfred Nobel's support of scientific research and after the institute where the work was done."[60]

However, the results obtained in Stockholm could not be reproduced by the Berkeley scientists and also not by Georgii Flerov and his collaborators at the Kurchatov Institute in Moscow, the predecessor of the JINR in Dubna. In experiments of 1958 Ghiorso, Seaborg, Torbjorn Sikkeland and John Walton announced to have positively identified the isotope No-254 and thus to have discovered the element.[61] A target of curium was bombarded with C-12 ions,

$$^{12}_{6}C + {}^{246}_{96}Cm \rightarrow {}^{254}_{102}Z + 4\,{}^{1}_{0}n,$$

and the nobelium isotope identified by its alpha decay. The Americans were confident that they had discovered the element and that the 1957 discovery claim was unfounded. The Stockholm group admitted that the Berkeley experiments "appeared to cast some doubt on our results" but concluded in a re-examination of early 1959 that the doubt was apparent only. Although the group maintained the validity of the 1957 investigation, "we suggest that judgment on the discovery of element 102 should be reserved."[62]

---

[57] Karol et al. (2016a).
[58] See Koppenol et al. (2016) for the most recent IUPAC naming criteria.
[59] Atterling et al. (1954).
[60] Fields et al. (1957). Accounts of the nobelium controversy, as seen from an American perspective, include Ghiorso and Sikkeland (1967) and Seaborg and Loveland (1990).
[61] Ghiorso et al. (1958).

The Russian physicists did not consider the American experiments to be more than just an indication of $Z = 102$, and in a series of experiments from 1958 to 1966 they provided definite proof of the existence of isotopes of the element. The Russians pointed to errors and inadequacies in the American experiments. On the other hand, the Americans disbelieved some of the Russian results or argued that they did not constitute a proper discovery. The Cold War atmosphere of disbelief between the two superpowers contributed to the controversy between the two groups.

As to the name of element 102, the Americans wanted to replace the too hastily assigned name nobelium with another one which better reflected the actual discovery. "Although the name nobelium for element 102 will undoubtedly have to be changed," wrote Seaborg in 1959, "the investigators have not, at the time of writing, made their suggestion for the new name."[63] However, nobelium had quickly come into common usage and entered textbooks and periodic tables, and for this reason the Americans refrained from proposing an alternative name. Also the Russian group asserted its priority and right to suggest a name. "Joliotium," referring to the French physicist, Nobel laureate and devoted communist Frédéric Joliot, was for a time the preferred name in the Soviet Union and later proposed also for the elements of atomic number 103 and 105.[64]

It took decades before the conflict over name and priority of the new element was finally settled. Only in reports from the early 1990s did IUPAC announce its decision. After a review of all relevant papers from the period 1957-1971 the TWG concluded in favour of Dubna experiments of 1966 which "give conclusive evidence that element 102 had been produced."[65] The TWG failed to find the same kind of evidence in the Berkeley work of 1958 which consequently was not accepted as the discovery of the element. The Berkeley group strongly disagreed, arguing that at least it should share the credit with the Russians. Seaborg found the TWG decision to be incredible and erroneous.[66]

---

[62] Fields et al. (1959).

[63] Seaborg (1959), p. 11.

[64] Joliotium was recommended for element 105 by IUPAC in 1994, but replaced by dubnium three years later. See Fontani, Costa and Orna (2015), pp. 385-388, and Rayner-Canham and Zheng (2008).

[65] Barber (1992).

The American specialists in SHE research protested vehemently and in general terms to the recommendations of the TWG concerning the elements of atomic numbers 101-112. Ghiorso and Seaborg plainly charged that the TWG panel was incompetent and their report "riddled with errors of omission and commission."[67] Among the complaints was that the panel, consisting of seven IUPAP members and only two IUPAC members, was biased toward physics with the consequence that the report was marked by a "downgrading of chemical contributions." Seaborg was trained in chemistry and considered himself a nuclear chemist and not a nuclear physicist. Since the Berkeley group relied much on expertise in nuclear chemistry, it thought that the alleged physics bias led to an unfair evaluation of the group's work. Paul Karol, a nuclear chemist at the Carnegie Mellon University, agreed. Calling the TWG study "flawed," he deplored that it included no experts in SHE chemistry.[68]

Moreover, Ghiorso and Seaborg complained that the TWG had had meetings with the Dubna scientists but none with the people from Berkeley. What Ghiorso and Seaborg did not comment on, but undoubtedly had in mind, was the national composition of the TWG. The nine members included one from Japan and eight Europeans, two of them from the former Soviet bloc. There were no Americans. Although the Soviet Union was dissolved by the end of 1991, the Cold War was not a forgotten chapter.

## 4. The transfermium wars

What has been called the "transfermium wars" refer to a series of disputes concerning the names and discoveries of primarily the elements with atomic numbers from 104 to 109 but also including a few other elements. The pugnacious name may first have been used by Karol, who in 1994 commented critically on the ongoing naming controversy.[69] A member of the ACS committee on nomenclature and later serving as chair of the JWP, Karol was centrally involved in this and other naming and priority controversies regarding the heavy synthetic elements.

---

[66] Seaborg (1994b), p. 242.

[67] Ghiorso and Seaborg (1993), which also appeared in Wilkinson et al. (1993b).

[68] Karol (1994). The only TWG member with experience in SHE research was the French physicist Marc Lefort. The German attitude to TWG was much more positive than the one of their American colleagues. Armbruster and Münzenberg (2012), p. 284.

## 4.1 A brief tour from $Z = 104$ to $Z = 109$

The earliest claim of having detected element 104, later known as rutherfordium (Rf), appeared in 1964 when researchers at JINR at Dubna reported to have produced the element when bombarding Pu-242 with Ne-22 ions.[70] The experiments did not clearly identify the atomic mass of the produced nuclide and were later shown to be partly invalid. Five years later a team of Berkeley scientists led by Ghiorso provided conclusive evidence by synthesizing $Z = 104$ from californium and carbon:

$$^{12}_{6}\text{C} + {}^{249}_{98}\text{Cf} \rightarrow {}^{257}_{104}\text{Rf} + 4{}^{1}_{0}n$$

Also element 105 (dubnium, Db) was first reported by Dubna scientists, who in 1968 and again in 1970 announced the identification based on an analysis of a nuclear reaction between Am-243 and Ne-22 ions. Later in 1970 the Ghiorso team reported experiments which questioned some of the Dubna results and independently resulted in a nuclide of element 105. By 1971, after further work by the Dubna group, the element was thought to have been discovered, but it was still unclear who had discovered it and if the discovery was conclusively established.

The case of $Z = 106$ (seaborgium, Sg) was even more controversial with regard to both discovery and name, and again it involved an extensive dispute between Berkeley and Dubna scientists.[71] The first evidence was reported in 1974 by a Dubna team led by Oganessian, the main evidence being events of spontaneous fission of what was supposed to be the nuclide of mass number 259. Shortly thereafter a team of scientists from Berkeley and LLNL, among them Ghiorso and Seaborg, produced the $A = 263$ nuclide. They identified the atomic number of the nuclide by the alpha decays of the new nuclide and its daughter nuclide:

$$^{263}_{106}\text{Sg} \rightarrow {}^{259}_{104}\text{Rf} + {}^{4}_{2}\text{He} \quad \text{and} \quad {}^{259}_{104}\text{Rf} \rightarrow {}^{255}_{102}\text{No} + {}^{4}_{2}\text{He}$$

---

[69] Karol (1994); Rothstein (1995).
[70] A summary of the discoveries and the relevant references appear in Seaborg and Loveland (1990), pp. 51-64.
[71] See Armbruster (1985) for a technical review of elements 106-109 as known at the time.

To proceed numerically, for once the Berkeley scientists were not centrally involved in the discovery of element 107 (bohrium, Bh). Once again Oganessian's Dubna group came first with a discovery claim dating from 1976 and which also included element 105. But as far as element 107 was concerned, the Russians were wrong as the spontaneous fission of Bh-261 on which they based their claim, was never confirmed. The element, in the form of the alpha-active nuclide with $A$ = 262, was only definitely detected in GSI experiments from 1981 led by Peter Armbruster and Gottfried Münzenberg.[72] These "cold fusion" experiments led the group to identify four events as decays of Bh-262.[73]

Also the next two elements were discovered in Darmstadt, although not without competition. What became known as hassium (Hs, $Z$ = 108) was discovered by the GSI team in 1984 but had a little earlier been claimed by Oganessian's group in Dubna. The delicacy of the GSI experiments are illustrated by the number of produced atoms sufficient for the identification of element 108: in one experiment three atoms of Hs-265 were produced and in another experiment just a single atom of Hs-264. The discovery claim of the Dubna group was tacitly withdrawn. Finally, element 109 (meitnerium, Mt) was identified by the GSI scientists a little earlier than element 108. The discovery event appeared in 1982 and a full paper on the new element was published six years later. The detected nuclide Mt-266 was alpha radioactive with a half-life of approximately 5 millisecond.

### 4.2. Criteria for discovery

The confusing number of discovery claims for new SHE through the 1960s and 1970s inevitably caused reconsideration of the old question, what does it mean to have discovered a new element? The answer to the question implied answers to priority claims and also, since it was generally agreed that the discoverers had the right to propose a name, related to the names of the new elements. CNIC addressed the problem, which in a minute from a meeting in 1969 was explained as follows:

---

[72] The GSI discoveries of elements 107-109 are described in Armbrüster and Münzenberg (2012).

[73] Early heavy-ion experiments producing SHE were "hot fusion" with the SHE nucleus in a highly excited state. The alternative "cold fusion" method of producing nuclei at low excitation energy was introduced by Oganessian in the 1970s and became the favourite method in Dubna and Darmstadt. For many years the method was regarded with suspicion in Berkeley. The SHE cold fusion has only the name in common with the sensational claim in 1989 of having produced hydrogen-to-helium fusion at room temperature.

> It had become abundantly clear during the past year that nuclear physicists reporting the preparations of new elements were exceedingly strongly attached to the right of the discoverer to select a name, and reluctant to accept the Commission's principle that whilst early selection of a name was a matter of convenience it carried no implications regarding priority of discovery.[74]

CNIC suggested that new elements should not be named until five years had elapsed after the first discovery claims. "The period would, it was hoped, allow confirmation of the initial discovery in another laboratory, and preferably in another country." American and Russian SHE scientists generally ignored the suggestion. With the purpose to consider the rival Berkeley-Dubna claims relating to elements 104 and 105, in 1974 IUPAC in collaboration with IUPAP appointed a group of nine supposedly neutral experts chaired by a British chemistry professor. However, the initiative was a failure as the committee never issued a report. Indeed, it never met as a group.[75] The more definite reasons for the failure are not known, but it seems that communication problems with the American and Russian laboratories were in part responsible.

In 1976 a group of Western SHE specialists, including Seaborg from Berkeley and Günter Hermann from Darmstadt, pointed out that lack of definite discovery criteria "has contributed significantly to the competing claims for the discovery for these [transuranic] elements."[76] In discussing various ways of identifying new elements, some chemical and other physical, the authors emphasized proof of the atomic number as essential. On a more practical note, "any claim to such a discovery should be published in a refereed journal with sufficient data to enable the reader to judge whether the evidence is consistent with such criteria." And as to the name of the new element, it "should not be proposed by the discoverers until the initial discovery is confirmed." This might have been a reference to the nobelium case.

Shortly before the establishment of the TWG, the leading GSI physicist Peter Armbruster wrote a review article in which he suggested a rule for the naming of SHE. At the same time he expressed his wish that the many controversies over names were taken care of by a commission of physicists and nuclear chemists. Armbruster's proposal, which inspired the not yet established TWG, was this:

> Element synthesis becomes production of a given isotope, and a name should be accepted only if the experiment claiming the discovery is reproducible. An isotope is defined by its mass and atomic number, its fingerprints are its decay modes and its half-life. … The proposed rule should be applied retrospectively for all elements discovered by isotope identification, that is elements 102-109.[77]

---

[74] Quoted in Fennell (1994), p. 269. See also Section 3.1.
[75] Wapstra et al. (1991), p. 881. Three of the committee members were Russians and three were Americans. According to Fennell (1994), p. 269, "In 1977 IUPAP said it had lost interest as the existence of the two elements was doubtful anyway."
[76] Harvey et al. (1976). The authors included seven Americans, one German and one Frenchman but none from the Soviet Union.

The rule was not primarily about names, but more about criteria for discovery. Its emphasis on reproducibility was reflected in the later TWG criteria, but in a less rigid formulation. Indeed, it is a problematic concept and Armbruster did not explain precisely what he had in mind when claiming that experiments must be reproducible.

In a report of 1992 the TWG investigated systematically and thoughtfully criteria for recognizing the existence of a new chemical element. The group fully realized the complexity of the discovery concept, implying that it could not be codified in a simple way. As shown by the history of element discoveries there had always been dissenting views of when and by whom an element was discovered, and the group argued that this was inevitable and had to be accepted:

> Different individuals or different groups may take different views as to the stage in the accumulation of evidence at which conviction is reached and may take different views as to the existence or otherwise of crucial steps leading to that conviction and as to which those crucial steps were. Such differences can be perfectly legitimate scientifically, in that they may depend upon, for example, differing views as to the reliability of the inference that might be drawn from certain types of evidence, while not disputing the reliability of the evidence itself. So, although the scientific community may reach consensus as to the existence of a new element, the reaching of that consensus is not necessarily a unique event and different views may, in all scientific honesty, be taken as to the steps by which it was reached.[78]

Nonetheless, the TWG came up with the following summary definition: "Discovery of a chemical element is the experimental demonstration, beyond reasonable doubt, of the existence of a nuclide with an atomic number Z not identified before, existing for at least $10^{-14}$ s." With regard to the requirement of a minimum lifetime of the nuclide it was introduced to make the formula more chemical and in accord with the standard view of the term element. It takes about $10^{-14}$ second for a nucleus to acquire its outer electrons and thus to become an atom with chemical properties. The same requirement was mentioned in the American-German-French 1976 proposal: "We suggest that composite nuclear systems that live less than about $10^{-14}$ second … shall not be considered a new element."[79]

---

[77] Armbruster (1985), p. 175. See also Armbruster and Münzenberg (2012), pp. 283-284.
[78] Wapstra et al. (1991), p. 882. See also Wilkinson et al. (1993a).

So-called quasi-atoms of very high $Z$ are formed transiently in heavy-ion collisions, but they exist only for about $10^{-20}$ second. Consequently they do not qualify as nuclides of new elements. However, there is no consensus among nuclear physicists of when a nucleus exists. Some take the definition of an atomic nucleus to be limited by the time scale $10^{-12}$ second, and according to others, "if a nucleus lives long compared to $10^{-22}$ s it should be considered a nucleus."[80]

The TWG realized that the phrase "beyond reasonable doubt" was vague, but the group used it deliberately to stress that such doubts could not be completely avoided. Confirmation of a discovery claim was of course important, but it was not a magic wand that in all cases would eliminate doubts. "All scientific data, other than those relating to unique events such as a supernova, must be susceptible of reproduction," it was pointed out.[81] And yet, although the TWG attached much importance to reproducibility, "We do not believe that recognition of the discovery of a new element should always be held up until the experiment or its equivalent have been repeated, desirable in principle as this may be." The more relaxed attitude to the reproducibility criterion was in part practically motivated in "the immense labour and the time necessary to detect even a single atom of a new element." There were circumstances, it was said, where "a repetition of the experiment would imply an unreasonable burden."

The TWG evaluation of discovery claims concerning the transfermium elements with atomic numbers from 101 to 112 appeared in 1992. The aim of the detailed report was to assign priority and credit for the discoveries of the elements, not to propose names for them.[82] This was done by reviewing all relevant papers and critically comparing their results with the discovery criteria and also with the most recent knowledge. In some cases the TWG conclusions were unambiguous, as they were in the case of $Z = 107$ where priority was assigned to the 1981 work of the Darmstadt group. But in other cases the conclusions were far from unambiguous. Element 103 provides an example. "Effective certainty" had been approached with papers published by the Dubna group in the late 1960s, but it was only with a Berkeley paper of 1971 that "all reasonable doubt had been expelled." So who should be credited with the discovery? The TWG tried to please both parties:

> In the complicated situation presented by element 103, with several papers of varying degrees of completeness and conviction, none conclusive, and referring to several isotopes, it is impossible to say other than that full confidence was built up over a decade with credit attaching to work in both Berkeley and Dubna.

---

[79] Harvey et al. (1976).
[80] See Thoenessen (2004), p. 1195.
[81] For the comparison to unique but accepted astronomical events in nature, see also Armbruster and Münzenberg (2012), p. 284. There are other kinds of unique and hence non-reproducible natural phenomena, for instance magnetic storms, cosmic rays and rare elementary particles which cannot be produced in the laboratory.

For elements 104 and 105 the TWG similarly suggested that credit should be shared between the American and Russian scientists, something which the Berkeley scientists found most unreasonable for $Z = 104$ in particular. For $Z = 106$, on the other hand, the verdict was clear and to the satisfaction of the Americans: "The Dubna work … does not demonstrate the formation of a new element with adequate conviction, whereas that from Berkeley-Livermore does." Priority for having discovered elements 107 to 109 was essentially, but in some cases with qualifications, assigned to the Darmstadt group. TWG refrained from describing element 109 as discovered and merely concluded that the work of the Darmstadt scientists "gives confidence that element 109 has been observed." Proof was still lacking.

The three SHE laboratories were given the right to respond to the TWG report.[83] While the Russian and German scientists were reasonably satisfied, the Americans were not. To their eyes, the transfermium war was not over. As mentioned above, Ghiorso and Seaborg reacted strongly against the conclusions and the TWG in general. Their "most serious quarrel" with the report concerned element 104 and the shared credit for its discovery to both Dubna and Berkeley.[84] Ghiorso and Seaborg flatly denied the validity of the TWG conclusion, which they considered "a 28disservice to the scientific community," and they also argued that priority to element 105 did not belong jointly to Berkeley and Dubna but to Berkeley alone. But Wilkinson, the chairman of the TWG, was not swayed: "After detailed examination of all criticism from Berkeley we do not find it necessary in any way to change the conclusions of our report."[85]

---

[82] Barber et al. (1992); Wilkinson et al. (1993a).
[83] Wilkinson et al. (1993b). Oganessian and I. Zvara responded on behalf of Dubna, and Münzenberg on behalf of Darmstadt.
[84] Ghiorso and Seaborg (1993).

The transfermium wars were as much about names as about priority. Recommendations for the names of elements 101-109 were agreed upon at a meeting of CNIC in the summer of 1994 and shortly later accepted by the IUPAC council (see the table).[86] The names were chosen after the three major SHE laboratories had been consulted, but to some SHE scientists they came as a surprise. The nomenclature committee of the American Chemical Society (ACS) had a few months earlier opted for rutherfordium instead of dubnium (104), hahnium instead of joliotium (105), and seaborgium instead of rutherfordium (106). Moreover, the Americans at first preferred nielsbohrium to bohrium, for other reasons because they found the latter name to be too close to boron. For element 108 they recommended the name chosen by the Darmstadt group, referring to the German state Hessen where GSI was located.

Not only did the Americans come up with names different from those of IUPAC, they also questioned the international union's authority to name elements and to disregard the proposals of the discoverers.[87] But IUPAC made it clear that it while it would consider such proposals it was not obliged to follow them. Due to persistent American pressure the 1994 list of names was not officially approved by the IUPAC council and at a meeting in Guilford, UK, in 1995 it was replaced by a different list of names thought to be a compromise acceptable to all parties. Only in 1997 was the final decision taken by the council. Dubnium was still on the list but now for element 105, flerovium had disappeared and so had joliotium.[88] While the discoverers of element 107 wanted to call it nielsbohrium, after consultations with the Danish IUPAC committee it was decided to retain the former bohrium.

One name that did not make it to the official lists but was advocated by the Dubna scientists, was kurchatovium for element 104 (and at one stage also for element 106). The name was originally proposed by Flerov in honour of Igor

---

[85] Wilkinson et al. (1993b), p. 1824. See also Bradley (1993).
[86] IUPAC (1994).
[87] Rothstein (1993); Rayner-Canham and Zheng (2008).
[88] IUPAC (1997). See also Fontani, Costa and Orna (2015), pp. 366-388.

Kurchatov, a nuclear physicist known as the father of the Soviet atomic bomb. Although kurchatovium went unrecognized in the West, it appeared in the 1979 edition of *The Great Soviet Encyclopedia*.[89]

| $Z$ | IUPAC 1970 | ACS 1994 | IUPAC 1994 | IUPAC 1997 |
|---|---|---|---|---|
| 102 | unnilbium Und | nobelium | nobelium | nobelium, No |
| 103 | unniltrium Unt | lawrencium | lawrencium | lawrencium, Lr |
| 104 | unnilquadium Unq | rutherfordium | dubnium | rutherfordium, Rf |
| 105 | unnilpentium Unp | hahnium | joliotium | dubnium, Db |
| 106 | unnilhexium Unh | seaborgium | rutherfordium | seaborgium, Sg |
| 107 | unnilseptium Uns | nielsbohrium | bohrium | bohrium, Bh |
| 108 | unniloctium Uno | hassium | hahnium | hassium, Hs |
| 109 | unnilnilium Une | meitnerium | meitnerium | meitnerium, Mt |

## 4.3. The case of seaborgium

According to the TWG reports of 1992 and 1993 the experiments made by the Berkeley-Livermore group in 1974 demonstrated the existence of element 106 "with adequate conviction." The nearly simultaneous work of the Dubna group was recognized to be important but not carrying the same conviction. While priority to the discovery of the element was thus assigned the American scientists, in its report of 1994 CNIC recommended a name for it (rutherfordium) which disregarded the proposals of the discoverers. Neither the Berkeley-Livermore group nor the rival Dubna group had suggested a name for the new element in their papers of 1974, the two groups agreeing that names should be postponed until their observations had been confirmed.[90]

Latest by 1994 the American claim for $Z$ = 106 was fully confirmed. Several names were proposed and discussed in Berkeley, some of them seriously and others more jokingly. Ghiorso suggested calling it "seaborgium" and after some hesitation 82-year-old Seaborg assented.[91] The proposal was announced at a meeting of the ACS in March 1994 and the name endorsed by the ACS committee on nomenclature. When the alternative CNIC recommendation became known half a year later, it met strong opposition from the powerful ACS and from the American scientific community generally. Ghiorso found the recommendation to be simply "outrageous."[92] And Karol deplored the "madness" of CNIC, which had rejected "for the first time in history, the name picked up by the undisputed discoverers of an element because the person so honoured was still alive." He had no confidence at all in what he called "the ongoing IUPAC transfermium fiasco."[93]

---

[89] For the richness of forgotten or alternative names of chemical elements, see Seaborg (1994b), Leal (2014), and Fontani, Costa and Orna (2015).

[90] Ghiorso et al. (1974), p. 1493. Members of the two groups had met in Berkeley and exchanged information about their experiments.

[91] Seaborg (1995).

Central to the short-lived dispute were two questions: (i) Do discoverers have the right to name an element? (ii) Can an element be named after a living person? With regard to the first question, IUPAC insisted that although discoverers have a right to *suggest* a name, it is IUPAC alone which makes the decision. As to the second question, in 1994 the twenty members of CNIC "resolved … that an element should not be named after a living person."[94] Not only was there no precedence for this, the commission also argued that it was necessary to have a proper historical perspective in relation to the discoveries of elements before the decision of a name could be made.

Seaborg protested: "In the case of element 106, this would be the first time in history that the acknowledged and uncontested discoverers of an element are denied the privilege of naming it."[95] Moreover, supporters of the Berkeley-ACS position argued that there were in fact historical precedents, namely in the cases of einsteinium and fermium. However, the argument bore no weight since the paper in which the names were proposed was published after the deaths of Fermi and Einstein (28 November 1954 and 18 April 1955, respectively) and the names approved by IUPAC only in 1957.[96] As *The Economist* generalized in a comment,

---

[92] Browne (1994).
[93] Karol (1994); Lehrman (1994).
[94] IUPAC (1994), p. 2420. The commission included members from 12 countries, three of them being American chemists.
[95] Quoted in Rayner-Canham and Zheng (2008), p. 17.
[96] Koppenol (2005). American scientists published several short papers in *Physical Review* in 1954 on the new elements 99 and 100, but without suggesting names. The discovery paper by Ghiorso and collaborators in the August issue of *Physical Review* was received on 20 June, two months after Einstein's death. It is possible, as suggested in Seaborg (1995b), that the Ghiorso group "decided on the names … while Albert Einstein and Enrico Fermi were still alive," but if so it was informally and hence of no relevance.

"When it comes to giving things names, scientists have a habit of throwing logic and consistency out of the window."[97]

IUPAC could have maintained its position, but in the end it did not. At a meeting in August 1996, CNIC "decided to modify its decision that the name of a living scientist should not be used as the basis for an element name."[98] The modification was ratified by the IUPAC council a year later. Seaborgium was the first element ever named after a scientist during his lifetime. Only in 2016, with the approval of oganesson for element 118, was another element named after a living scientist, in this case to Yuri Oganessian, another octogenerian.

The question of the names of artificially produced elements was not a new one. In 1947, at a time when eight such elements were known, the eminent Austrian-British radiochemist Friedrich Paneth addressed the question. He suggested three rules that he thought would lead to greater unity and consistency in the chemical names:

> (1) The right to name an element should go to the first to give a definite proof of the existence of one of its isotopes. (2) In deciding the priority of the discovery, there should be no discrimination between naturally occurring and artificially produced isotopes. (3) If a claim to such a discovery has been accepted in the past, but is refuted by later research, the name given should be deleted and replaced by one chosen by the real discoverer.[99]

The second of the suggestions was at the time uncontroversial, but the first one was not and was in fact rejected by IUPAC at its London conference the same year. Nor did IUPAC or the majority of scientists accept the third suggestion, which too easily might lead to confusion and frequent changes of names. Still, during the period of the transfermium wars this is what happened with several of the names assigned to the superheavy elements. The procedure suggested by Paneth was not ineffective, though, as the revision of element names adopted by IUC in 1949 was directly inspired by Paneth's rules.[100]

---

[97] *The Economist*, December 1998. https://www.astro.com/swisseph/econ4686.htm
[98] IUPAC (1997), p. 2472.
[99] Paneth (1947), p. 8; Koppenol (2005).
[100] Coryell and Sugarman (1950).

## 5. Super-superheavy elements

The elements with atomic numbers 110, 111 and 112 were produced, almost routinely, by the Darmstadt scientists in the brief period 1995-1996.[101] Only in the case of $Z = 110$ did they face serious competition, in this case from the Dubna group. But although the search for the element was competitive, the relationship between the two groups was friendly and cooperative. The first two elements received names and official recognition relatively quickly – or by IUPAC standards very quickly. CNIC recommended darmstadtium (Ds, 110) as a new element in 2003 and roentgenium (Rg, 111) was recommended the following year. The last of the trio, copernicium (Cn, 112) had to wait until 2009, after earlier assessments of the JWG had concluded that the evidence for the discovery claim was still insufficient.[102]

Elements 114 and 116 were Dubna discoveries made by a team led by Oganessian but with participation of American researchers from LLNL, an indication that the transfermium war was over. The early history of $Z = 114$ included several unconfirmed and doubtful events ascribed to decay of just one or two atomic nuclei. Only after the nuclides with mass numbers between 286 and 289 had been confirmed by experiments in Berkeley and Darmstadt was the JWG satisfied and concluded in its report of 2011 that the element had been discovered: "The JWP … recommends that the Dubna-Livermore collaboration be credited with the discovery of this new element. In a similar manner, … the Dubna-Livermore collaboration should be credited with the discovery of the new element with $Z = 116$."[103] After the false discovery claim of 1999 (Section 3.3) attempts to produce element 116 were continued by the Dubna and Livermore scientists and it was their work which eventually led the JWP to accept that the element had been discovered.

The names of the two elements did not involve controversy. On the suggestion of the Dubna team element 114 was named flerovium, not after Georgii Flerov personally but after the Flerov Laboratory of Nuclear Reactions, a part of the JINR. The same kind of institutional naming was used for element 116, livermorium, the name being a reference to the LLNL in Livermore, California. Both names were approved by the IUPAC council in 2012.[104]

---

[101] Hofmann (2002), pp. 163-174; Armbruster and Münzenberg (2012), pp. 288-300.
[102] Karol et al. (2001); Karol et al. (2003); Barber et al. (2009).
[103] Barber et al. (2011), p. 1494.
[104] Loss and Corish (2012).

| Z   | name, symbol      | discovery | group            | JWG recognition | IUPAC recognition |
|-----|-------------------|-----------|------------------|-----------------|-------------------|
| 110 | darmstadtium, Ds  | 1995      | Darmstadt        | 2001            | 2003              |
| 111 | roentgenium, Rg   | 1995      | Darmstadt        | 2001            | 2004              |
| 112 | copernicium, Cn   | 1996      | Darmstadt        | 2009            | 2009              |
| 114 | flerovium, Fl     | 1999      | Dubna/Livermore  | 2011            | 2012              |
| 116 | livermorium, Lv   | 2004      | Dubna/Livermore  | 2011            | 2012              |

The most recent newcomers to the periodic table are the elements with atomic numbers 113, 115, 117 and 118. The press release of IUPAC on 30 December 2015 caused public attention and several speculations of what names would be assigned to the new elements. It also caused some consternation, as apparently it was due to IUPAC alone, without consulting IUPAP.[105] Following a five-month period of public review, the names of the new elements were approved by the IUPAC bureau on 28 November 2016. They were ratified by the IUPAC council at a meeting in July 2017.

Disregarding an earlier discovery claim by the Dubna-Livermore collaboration, the first atoms of element 113 were produced in 2003-2005 by a team led by Kōsuke Morita at the RIKEN Nishina Center for Accelarator-Based Science (RNC) in Japan. The name refers to Yoshia Nishina, a pioneer Japanese physicist and former collaborator of Niels Bohr. While previously SHE research had been limited to institutions in USA, Russia and Germany, now the RIKEN group entered the game as an important fourth player. Although the JWP did not accept the RIKEN claim as a proved discovery in its report of 2011, after further work it did so five years later.[106] The new element, the first one ever discovered in Asia and by scientists from this continent, was named nihonium after the country where it was discovered.[107]

In experiments of 2004 Oganessian's Dubna-Livermore collaboration not only claimed the discovery of $Z = 113$ but also of $Z = 115$ as the alpha parent element of the former. Whereas priority to element 113 was credited the RIKEN group, after further experiments and confirmations from other groups the JWP recognized in its 2016 report the Dubna-Livermore collaboration as the discoverer of element 115 and also of element 117. Indeed, the discovery histories of the two elements were closely intertwined. The experiments leading to the discovery claim of element 117 in 2010 involved not only the Dubna and Livermore groups but also scientists from the Oak Ridge National Laboratory in Tennessee. Scientists from the three groups agreed to call it tennesine (after Tennessee) and IUPAC followed the recommendation. As to element 115 it was named moscovium (after Moscow), thus diplomatically honouring both an American and a Russian locality. As illustrated by the discovery paper in which element 117 was announced, SHE production was big science. The paper by Oganessian and his collaborators included 33 authors from six different institutions.[108]

---

[105] Burdette et al. (2016); Jarlskog (2016).
[106] Karol et al. (2016b).
[107] Öhrström and Reedijk (2016).

And finally to element 118 which was another product of the successful Dubna-Livermore collaboration. Its discovery history was summarized in Section 3.3 and here we just note that so far no more than a handful of nuclei have been produced. The only known isotope is Og-294, which makes it the element with the fewest isotopes currently known. The nuclide decays with a lifetime less than 1 millisecond. None of the physical and chemical properties of the element have been measured. They are all calculated or extrapolated, including the number and structure of the element's atomic electrons. Of course, no chemical compounds of oganesson are known.

## 6. Philosophical issues

### 6.1. Between physics and chemistry

There is no doubt that SHE research is basically nuclear physics and has been so since the synthesis of the first transuranic elements. In the later development SHE experiments and interpretation of data have relied crucially on advanced detectors from high energy physics that are foreign to the research tradition of chemistry. What is more, contributions to SHE science are predominantly published in physics journals (such as *Physical Review Letters* and *European Physical Journal*) and not in chemistry journals.

Nonetheless, SHE is about elements and there is a long historical tradition that everything concerning new elements belong to the domain of chemistry. Radioactivity is a property of certain elements which transmute to other elements and this was the rationale for counting much of early research in radioactivity to chemistry rather than physics. Rutherford did not hold chemistry in high esteem and yet his 1908 Nobel Prize was in chemistry, not physics. Nuclear chemistry, the successor of early radiochemistry, remained a chemical discipline. The responsibility of recognizing new elements belonged to IUPAC or its predecessors, and although the interdisciplinary working groups TWG and JWP included physicists, the final decision rested and still rests with IUPAC. Its sister union IUPAP stayed on the sideline. The distinction between physics and chemistry in SHE research is in some way artificial as workers in the field rarely consider themselves as either physicists or chemists. Still, the relationship between the two sister sciences and their respective organizations is not irrelevant and has not always been harmonious.

---

[108] Oganessian et al. (2010).

As mentioned in Section 3.4, in the early 1990s American SHE researchers including Seaborg, Ghiorso and Karol complained that the TWG panel was dominated by physicists who did not appreciate the methods of nuclear chemistry. While previous disagreements between chemists and physicists were rarely openly discussed, recently the subject has been addressed in a remarkably candid way, but this time from the physics point of view. The Swedish theoretical physicist Cecilia Jarlskog had been a member of the Nobel Prize physics committee and she served from 2011 to 2014 as president of IUPAP. In an address given to a Nobel Symposium of 2016 she declared war against the IUPAC managerial staff which she accused of being incompetent in matters concerning SHE. Not only was IUPAC responsible for the "failure" of the most recent JWP, it had also stolen the credit from the physicists. And that was not all:

> Everyone agrees that the discovery of the new superheavy elements, a process which takes ages, lies in the 'camp of physicists'. … Our physicists deserve and need to be publicly acknowledged for their achievements… [but] the IUPAC managerial staff has gone behind our back and violated the 'ethical rules', in spite of not having the competence to validate the discoveries."[109]

Jarlskog consequently suggested that IUPAP and not IUPAC should take the leading role in the SHE enterprise. Only the future can tell whether this will happen or not. When Jarlskog entered the negotiations with IUPAC to establish a new working group, she "could not imagine that there would be 'political' aspects in it." She admitted being "naïve and believing that scientists are impartial and logical."

---

[109] Jarlskog (2016).

## 6.2. The discovery of an element

To discover some object X is to observe for the first time that X exists and to demonstrate the observation claim convincingly, to the satisfaction of the relevant scientific community. (I shall here disregard the discovery concept in relation to phenomena, laws and theories.) It follows that it is not sufficient to postulate or hypothesize that X exists, as this will not satisfy the community. Moreover, the discoverer must not only have observed X but also recognize that X is a novel object or phenomenon with certain properties. The object in question obviously has to exist at the time of the discovery, but not necessarily at an earlier or later time. Or perhaps one should say, more cautiously, that people at the time of the discovery have to believe that the object exists.[110] Discoveries are not necessarily stable events or assignable to a definite discoverer. A novel observation of X made by one person may later be made by another person in ignorance of the previous person's work, in which case we say that X was re-discovered. There are also examples of what may be called de-discoveries, namely that what was accepted as a discovery was later robbed this status and declared a non-discovery.

      The concept of discovery plays a central role in the reward system of science but is rarely defined or problematized. For example, Nobel prizes in science are awarded for discoveries and yet the Nobel Foundation has never formulated criteria for discovery similar to those found in the TWG reports of the early 1990s. The authors of these reports were fully aware of the complexity of the discovery concept, which they discussed with no less sophistication than textbooks in philosophy of science. Here is an example:

> A discovery is not always a single, simply identifiable event or even the culmination of a series of researches in a single institution, but may rather be the product of several series of investigations, perhaps in several institutions, perhaps over several years, that has cumulatively brought the scientific community to the belief that the formation of a new element has indeed been established. However, since different sections of the scientific community may have different views as to the importance and reliability of interpretation of different sorts of scientific evidence, the bringing into that belief of these different sections of the community may well occur at different times and at different stages of the accumulation of the evidence. Where, then, does discovery lie?[111]

---

[110] Achinstein (2001), p. 405 emphasizes that what does not exist cannot be discovered. The claim is not without problems as it presupposes that our current understanding of what exists is permanent. From a historical perspective it makes sense to say that chemists discovered phlogiston around 1730 or that physicists around 1910 discovered that the atomic nucleus consisted of protons and electrons.

It follows that at least in some cases the notion of absolute priority is unrealistic. Those who demand absolute priority to be assigned in all cases cling to "outmoded concepts of the nature of discovery."[112]

Whatever the kind of object, a claim does not constitute a discovery before it has won broad recognition, which typically means that it has been independently confirmed by other scientists and accepted by the scientific community. For this reason the discovery claim and the reasons for it have to be announced in public, which normally happens in the form of a scientific paper but can also be in an oral presentation. The inevitable social dimension of the discovery concept can be boiled down to the statement that only those who can convince their scientific peers are credited with having made a discovery. This does not imply that the epistemic dimension is irrelevant. According to a social-constructivist view, "discoveries are … socially defined and recognized productions … [and they] occur because they are made to occur socially by processes of social recognition."[113] But although consensus is indeed a necessary element, it goes too far to claim that it is also sufficient and independent of whether the discovery claim is epistemically justified or not.

Discovery is not a discrete event which occurs almost instantaneously. It is a cognitive as well as social process that involves recognition from other scientists and their institutions. William Ramsay discovered the new gaseous element helium in the spring of 1895, but even in this relatively uncomplicated case it is difficult to identify a definite moment of discovery.[114] When Ramsay on 22 March identified new spectral lines in a gas evolved from the mineral cleveite, he realized that there was something new in the gas but only after two weeks of further work did he conclude that it was a new element. He then communicated his conclusion or discovery claim in private letters and on 25 April he announced it in a brief paper. The discovery claim was criticized by a few chemists but soon it was confirmed and generally accepted. Ramsay undoubtedly discovered a new element in the spring of 1895.

---

[111] Wilkinson et al. (1993a), p. 1759.
[112] Wapstra et al. (1991), p. 883.
[113] Brannigan (1981), p. 77 and p. 169.
[114] See Kragh (2009) for details.

The discovery of element 72 by George Hevesy and Dirk Coster in early 1923 was more controversial because it was part of a long and bitter priority conflict which involved the very nature of the element.[115] Was it a zirconium homologue, hence hafnium, or a rare earth metal, hence celtium? The element was the first one identified by means of its atomic number as revealed by its characteristic X-ray spectrum. Although by 1925 hafnium was accepted by the large majority of scientists, it still missed the official discovery stamp from the International Committee on Chemical Elements (ICCE), a branch under IUPAC and with several members sympathetic to the rival celtium. On the other hand, hafnium was quickly recognized by the German Atomic Weight Commission. In an attempt to secure neutrality, ICCE simply omitted element 72 from its 1925 version of the periodic table! The international committee chose a compromise in the politically sensitive question, namely by accepting two names and two symbols (Hf, Ct) for the same element.[116] Only in 1931, at a time when celtium survived in French chemistry only, did a new Committee on Atomic Weights approve hafnium as the only name for element 72.[117] Clearly, there were predecessors to the later discovery and naming disputes concerning SHE.

The story of SHE illustrates how the concept of discovery was turned into an operational concept in a particular area of science and how scientists were forced to reflect on the meaning of discovery. The definition of an element did not change as the atomic number $Z$ was still the defining property, but the TWG report of 1991 pointed out that "The exact value of $Z$ need not be determined, only that it is different from all $Z$-values observed before, beyond reasonable doubt." That determination of the atomic number was still important is shown by the competing claims for having found element 113. When the JWP decided to attribute the discovery to the RIKEN team and not to the Dubna team, it was primarily because the first team provided solid evidence for the atomic number. The Dubna

---

[115] Kragh (1980). See Kragh (2000b) for the change in the conception of a chemical element ca. 1910-1925.

[116] As the committee tersely stated, "The table does not include No. 72, Hafnium or Celtium, atomic weight 180.8." ICCE (1925), p. 597.

[117] Baxter et al. (1931). The International Union of Chemistry confirmed in 1949, that hafnium and not celtium was the official name. Coryell and Sugarman (1950).

39measurements, on the other hand, "were not able to within reasonable doubt determine Z."[118]

The 1991 report was not only an attempt to define SHE discovery operationally, it also included a discussion of discovery in more general terms (Section 4.2).[119] The criteria proposed by Wilkinson and his colleagues were in part conventional in so far that they referred to the public domain and to the importance of confirmation. But the report was also unconventional in the sense that it admitted that there were different legitimate views regarding "the accumulation of evidence at which conviction is reached." The evidence had to be solid and reliable, of course, but the interpretation of it could well be disputed.

Confirmation is only meaningful if experimental results are reproducible, but the TWG realized that although reproducibility is an important desideratum it is not always necessary. If a SHE experiment resulted in good data based on unobjectionable methods, confirmation could be dispensed with. This was a situation known from other parts of the physical sciences, such as astrophysics and high energy physics, where complicated and very expensive experiments could not be easily replicated. The complex nature of the discovery concept is further illustrated by TWG's and JWP's description of SHE discoveries as cumulative processes. Contributions from one laboratory to measurements obtained by another laboratory might eventually increase the confidence of the results to such a level that they constitute convincing evidence for the discovery of a new element. In such a case, who should be credited as the discoverer? As we saw in connection with the elements from $Z = 103$ to $Z = 105$, the TWG decided to share the credit between two competing teams, one from Dubna and the other from Berkeley and Livermore. The question of who discovered the elements and when cannot be answered with a name and a date.

### 6.3. Does oganesson exist?

Together with a few sub-uranium elements, the transuranic elements do not exist in nature but are artificially produced in laboratories. They nonetheless count as chemical elements inhabiting their proper places in the periodic table. We say that fermium was *discovered* in 1955 and meitnerium in 1982, but clearly the elements were discovered in a different sense than gallium and hafnium were discovered. SHE and artificial elements generally were *created* or *invented,* in largely the same way that a statue is created or a technological device invented. They belong to what the ancient Greeks called *techne* (human-made objects or imitation of nature) and not to *physis* (nature). The creation of synthetic and yet in a sense natural objects did not start with the work of Segré and Perrier in 1937, for at that time there already was a long tradition in organic chemistry of synthesizing chemical compounds. The first such compound without a counterpart in nature may have been William Perkin's famous discovery (or manufacture) of the aniline dye mauveine in 1856.

---

[118] Karol et al. (2016b), p. 146.
[119] Wapstra et al. (1991), reprinted in Barber et al. (1992).

Whereas plutonium may be said to be a technological product in so far that technologies are always purposeful and oriented towards social practices, this is not the case with most SHE. They have been produced in minute amounts only and serve no social or economic purposes. More importantly, they have short life-times and thus in several cases have been produced only to disappear again almost instantly. The half-lives of the longest-lived transfermium isotopes vary greatly and generally decrease with the atomic number, from 51.5 days (Md-258) to 0.7 milliseconds (Og-294). The elements have been produced and detected in nuclear processes and thus *did exist* at the time of the experiments. But strictly speaking they *do not exist* presently. Their existence is ephemeral or perhaps potential, which is quite different from the existence of ordinary elements whether radioactive or not. Can we truly say that the element oganesson *exists* when there is not a single atom of it in the entire universe? Sure, more atoms or rather nuclei of element 118 could be produced by replicating or modifying the Dubna experiments, but within a second they would disappear again.

One may object that particles with even shorter lifetimes are known from high energy physics without physicists doubting that they really exist. For example, the neutral pion decays into two gamma quanta with a lifetime of about $10^{-16}$ seconds. The particle was first detected in nuclear reactions in Berkeley in 1950, but contrary to nuclides of SHE it was also found in nature, namely in cosmic rays. The neutral pion exists and is not exclusively a laboratory product. The same is the case with the antiproton, another exotic particle first produced in accelerator experiments and only subsequently detected in the cosmic rays. The antiproton can be brought to combine with a positron and thus form anti-hydrogen, an exotic atomic system which has

41been produced in the laboratory and studied experimentally.[120] Anti-hydrogen has in common with SHE that it is element-like and only exists when manufactured. But contrary to SHE, there is no place for anti-hydrogen or other anti-elements in the periodic system.

A related question is whether short-lived transfermium elements really count as *elements* in the traditional meaning of the term. Elements consist of atoms and it is the atoms and their combinations which endow elements with chemical properties. An isolated atomic nucleus has no chemistry. Not only is the number of produced transfermium atoms extremely small, what is formed are nuclei and not atoms. Under normal circumstances a bare atomic nucleus will attract electrons and form an atom, but the circumstances of SHE experiments are not normal and atoms may not be formed at all or only exist for such a small time that they cannot be examined.

No atoms are known for the SHE with $Z > 109$ of which only atomic nuclei have been produced. Scientists nonetheless know much about these still hypothetical atoms, and about the physical and chemical properties of the elements, but the knowledge is exclusively in the form of theoretical predictions, extrapolations and estimates. For example, the electron structure and radius of the tennesine atom, element 117, have been calculated and so have the boiling point and density of the element as well as hypothetical compounds such as TsH and $TsF_3$. But there are no empirical data and none are expected to come as far as I know. The situation is different for the less heavy transuranic elements and especially for plutonium. Even some of the transactinides such as rutherfordium and dubnium have a real chemistry.[121]

Consider again the heaviest of the elements, oganesson, which is presently known only as one nuclide with an extremely small lifetime. Very few of the nuclei have been produced and none of them exist any longer. Oganesson has received official recognition and entered the periodic table alongside other elements and yet one may sensibly ask if oganesson is really a chemical element in the ordinary sense of the term. Perhaps its proper status is better characterized as a potential element.

---

[120] Amoretti et al. (2002).
[121] Kratz (2011).

44Karol, Paul J. et al. (2016b). "Discovery of the elements with atomic numbers $Z$ = 113, 115 and 117." *Pure and Applied Chemistry* **88**: 139-153.

Koppenol, Willem H. (2005). "Paneth, IUPAC, and the naming of elements." *Helvetica Chimica Acta* **88**: 95-99.

Koppenol, Willem H. et al. (2002). "Naming of new elements." *Pure and Applied Chemistry* **74**: 787-791.

Koppenol, Willem H. et al. (2016). "How to name new elements." *Pure and Applied Chemistry* **88**: 401-405.

Kostecka, Keith (2008). "Americium – from discovery to the smoke detector and beyond." *Bulletin for the History of Chemistry* **33**: 89-93.

Kragh, Helge (1980). "Anatomy of a priority conflict: The case of element 72." *Centaurus* **23**: 275-301.

Kragh, Helge (1999). *Quantum Generations: A History of Physics in the Twentieth Century*. Princeton: Princeton University Press.

Kragh, Helge (2000a). "An unlikely connection: Geochemistry and nuclear structure." *Physics in Perspective* **2**: 381-397.

Kragh, Helge (2000b). "Conceptual changes in chemistry: The notion of a chemical element." *Studies in History and Philosophy of Modern Physics* **31**: 435-450.

Kragh, Helge (2009). "The solar element: A reconsideration of helium's early history." *Annals of Science* **66**: 157-182.

Kragh, Helge (2013). "Superheavy elements and the upper limit of the periodic table: Early speculations." *European Physical Journal H* **38**: 411-431.

Kratz, J. V. (2011). "Chemistry of transactinides." In *Handbook of Nuclear Chemistry*, eds. A. Vértes et al., pp. 925-1004. Berlin: Springer.

Leal, João (2014). "The forgotten names of chemical elements." *Foundations of Science* **19**: 175-183.

Lehrman, Sally (1994). "'Seaborgium' fails to win approval." *Nature* **371**: 639.

Loss, Robert D. and John Corish (2012). "Names and symbols of the elements with atomic numbers 114 and 116." *Pure and Applied Chemistry* **84**: 1669-1672.

MacMillan, Edwin M. (1951). "The transuranium elements: early history." http://www.nobelprize.org/nobel_prizes/chemistry/laureates/1951/mcmillan-lecture.pdf

Marinov, Amnon et al. (1971a). "Evidence for the possible existence of a superheavy element with atomic number 112." *Nature* **229**: 464-467.

Marinov, Amnon et al. (1971b). "Spontaneous fission previously observed in a mercury source." *Nature* **234**: 212-215.

Marinov, Amnon et al. (1984). "Consistent interpretation of the secondary-reaction experiments in W targets and prospects for production of superheavy elements in ordinary heavy-ion reactions." *Physical Review Letters* **52**: 2209-2212.

Marinov, Amnon et al. (2004). "Response to the IUPAC/IUPAP Joint Working Party second report." Arxiv:nucl-ex/0411017.

Marinov, Amnon et al. (2010). "Evidence for the possible existence of a long-lived superheavy nucleus with atomic mass number $A$ = 292 and atomic number $Z \cong 122$ in natural Th." *International Journal of Modern Physics E* **19**: 131-140.